\documentclass[12pt,a4paper]{article}
 
\usepackage{graphicx}
\usepackage{caption}
\usepackage{subcaption} 

\usepackage[utf8]{inputenc}
\usepackage[english]{babel}
\usepackage{amsmath}
\usepackage{amsfonts}
\usepackage{amssymb}
\usepackage{makeidx}
\usepackage{graphicx}
\usepackage{lmodern}
\usepackage[left=2cm,right=2cm,top=2cm,bottom=2cm]{geometry}

\newcommand{\onehalf}{\frac{1}{2}}
\newcommand{\dis}{\displaystyle}

\newcommand{\paren}[1]{\left( #1 \right)}

\def\eg{{\it e.g.\ }}
\def\etal{{\it et al.\ }}
\def\ie{{\it i.e.\ }}

\begin{document}

\title{Understanding the physics of world's fastest non-motorized sport \\[.2cm] }

\author{\it \small Narit Pidokrajt\footnote{narit.pidokrajt@his.se} \\ \small \sl Division of Mechanical Engineering, Mathematics and Physics (MMF) \\ \small \sl School of Engineering Science, University of Skövde \\
 \small \sl 541 28  Skövde, Sweden.}

\date{\today}

\maketitle

\begin{abstract}

The main objective of this article is to describe the physics of {\bf Speed Skydiving}, world's fastest non-motorized sport practiced under FAI\footnote{FAI stands for Fédération Aéronautique Internationale (World Air Sports Federation), a recognized body by the International Olympic Committee.},  with the help of standard physics and mathematics taught during the first or second year for undergraduate students in physics worldwide. Although this discipline of air sports has been around since the middle of the 1990s and has been a part of major international sport events, it is not yet well-known to the wider public and has not been addressed thoroughly in physics textbooks as the physics of various other sports. 

It is hoped that this paper will introduce and engage physics teachers/instructors new to this air sport discipline---pedagogically.  University or high school students may find this article helpful while studying a subject of terminal speeds, which is thoroughly discussed herein. This article can also serve as a background for further studies in physics of skydiving and aerodynamics.
\end{abstract}

\noindent {\tt \small Keywords: Physics Education, Aerodynamics, Sport Science, Speed Skydiving, \\ Drag coefficients, Air sports, ISSA, ISC, FAI.}

\maketitle

\section{Introduction}

\begin{figure}
\begin{center}
\includegraphics[scale=0.1]{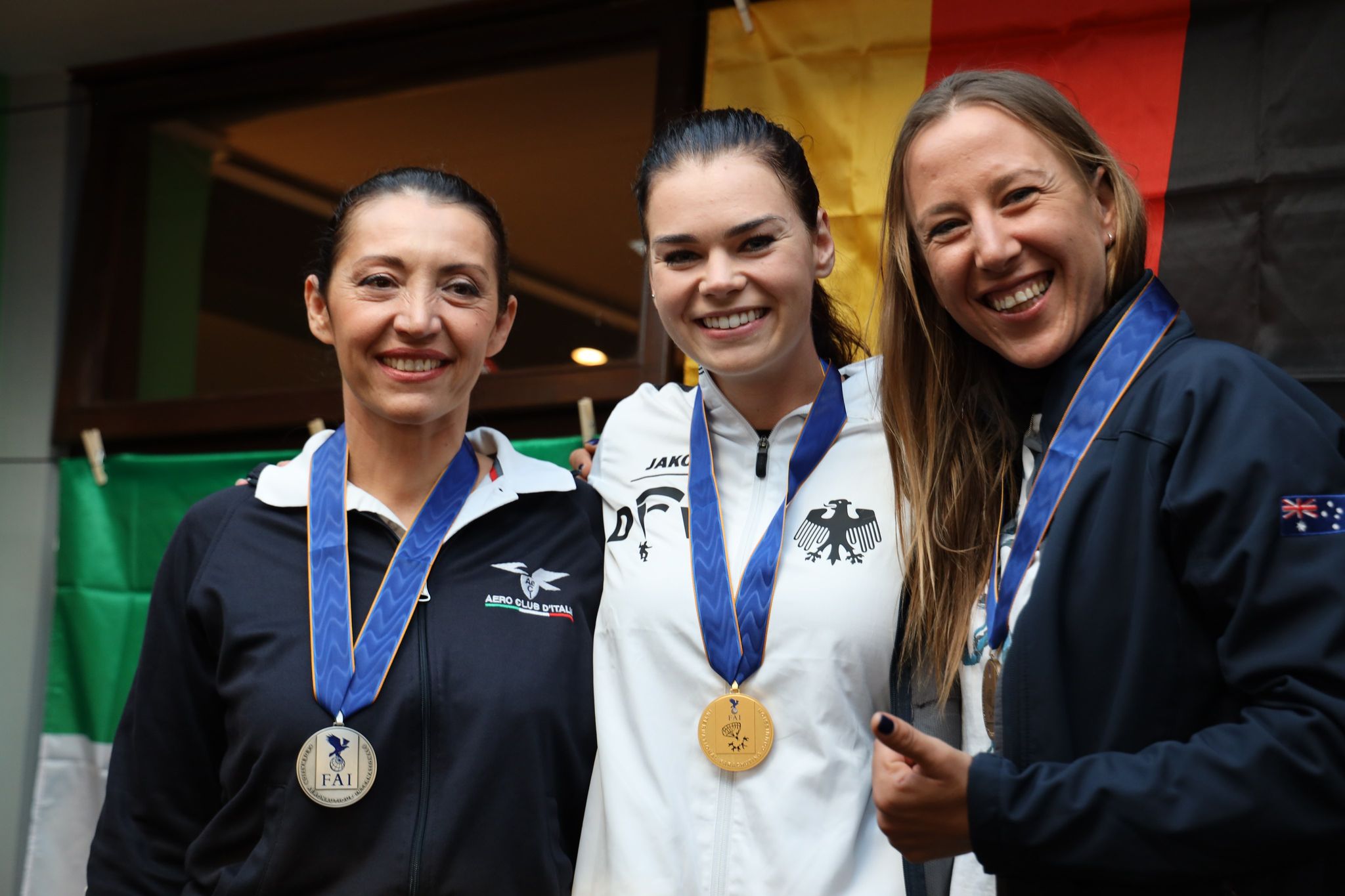} 
\caption{\sl Speed skydivers at a European championship award ceremony. The winner was Ms. Lucia Lippold (middle) who is a university graduate student, a member of the German National Team for Skydiving. Her record in speed skydiving is 417.68 km/h as measured by a GPS~\cite{paralog}.  Photo: Private collection of Lucia Lippold.}
\label{fig:Lucy}
\end{center} 
\end{figure}

Speed Skydiving is the fastest non-motorized sport where Earth's gravity and geometry of the skydiver through Earth's atmosphere play a major role in determining the outcome of the jump. It is a {\it space age} sport which relies heavily on sciences in almost all aspects of it, from the way the jump is performed to the equipment and the rules of the sport themselves.  The main goal of the athletes (speed skydivers) is to achieve the fastest freefall speed possible over a given distance.  There are in principle only two factors that influence the outcome of the jump according to physics: skydiver's mass and body orientation through the atmospheric friction (air resistance).  In short what counts is the terminal speed achieved through aerodynamics. How the athletes achieve their higest terminal speed during a jump can depend on a number of factors\footnote{\eg having a bad exit from the aircraft can affect the process of achieving the terminal speed negatively.}.

\noindent Speed skydiving is a discipline within air sports which heavily relies on technology for measurements. This skydiving discipline has been a part of international parachuting competitions for the past two decades. Rules and regulations have been updated a number of times.   The speed skydivers strive to achieve their higest terminal speed using the aforementioned aspects---whether or not they are  aware of the underlying physics of the sport.  
 
\begin{figure}[h!]
\begin{center}
\includegraphics[scale=0.25]{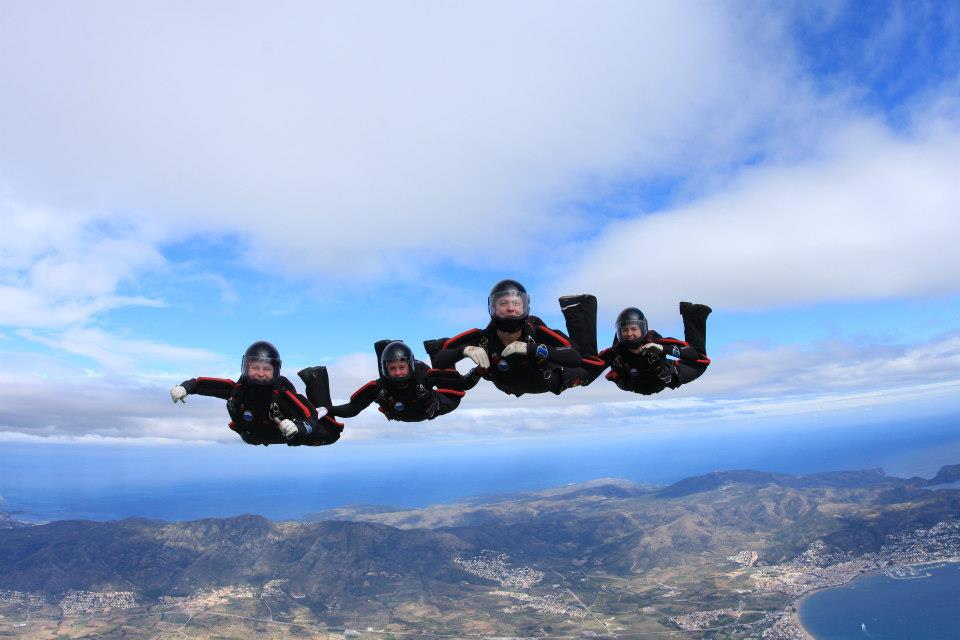} 
\caption{\sl Team Fire, a former national team of Sweden in 4-way skydiving formation (FS), under practice. In FS the skydivers face down toward the Earth with their belly at all times and hence their cross-sectional areas are always greater than their peers performing a so-called head-down body position whose $C_d$ can be approximated by a half-sphere. Photo: Team Fire.}
\label{fig:Teamfire}
\end{center} 
\end{figure}

\begin{figure}
\begin{center}
\includegraphics[scale=0.2]{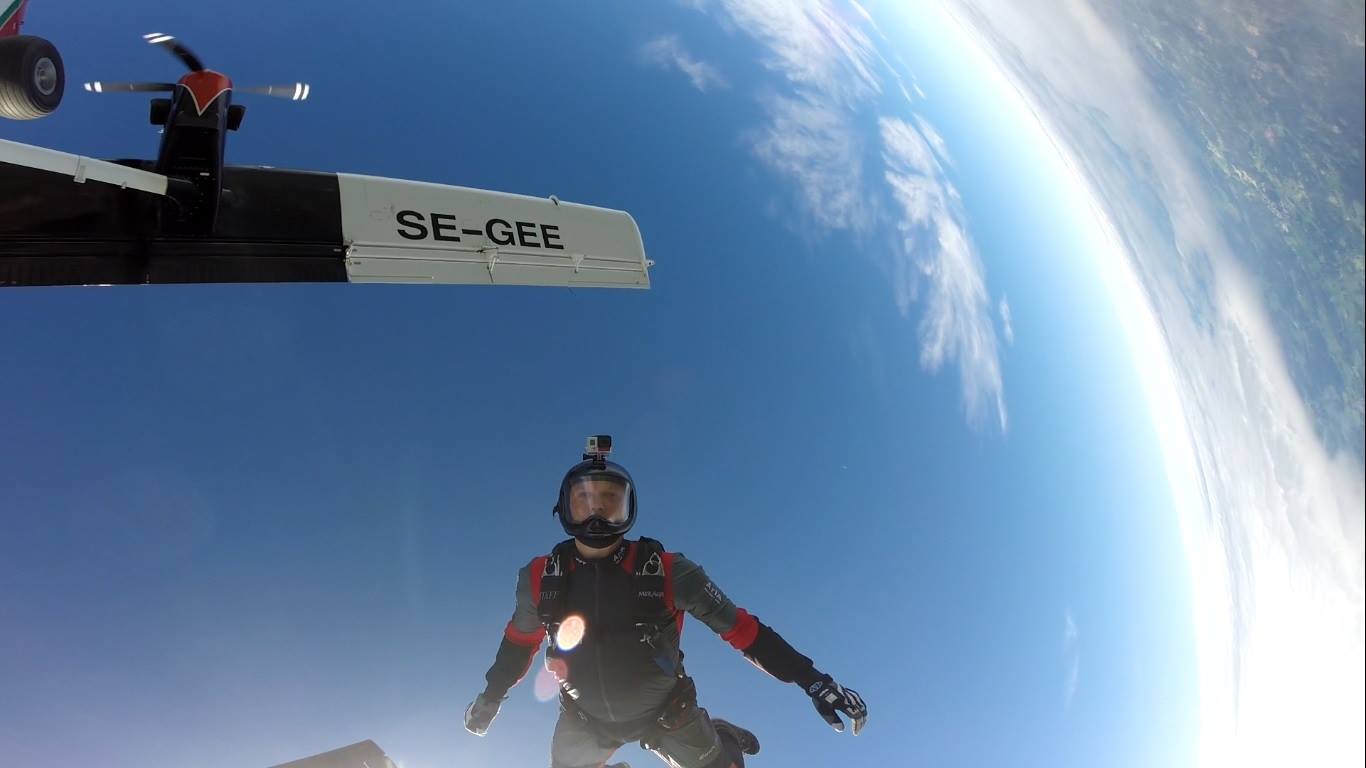} 
\caption{\sl A speed skydiver exiting an aircraft attempting to be as streamlined as possible. Shown in the figure is Mr. Daniel Hagström of Sweden's National Team right after an exit from the aircraft. He dives into his streamlined body orientation in order to fall to reach the maximum speed. Mr. Hagström's personal best is in excess of 470 km/h. Photo: Private collection of Daniel Hagström.}
\label{fig:Daniel}
\end{center} 
\end{figure}

In contrary to the standard form of skydiving in which skydivers face the Earth's surface in a belly position\footnote{In other freefall disciplines of the sport such as freefall style or free fly, skydivers are not necessarily in a belly position.} (Fig.~\ref{fig:Teamfire}) reaching typically 200 km/h, speed skydivers will try to form their body to be as streamlined as possible (see Fig.~\ref{fig:Daniel}). As such they easily reach speeds in the vicinity of over 400 km/h!

\noindent Skydiving itself captures the imagination of the public by opening a new world of adventure, speeds and unique experience of free-falling. Luckily this is something in which anyone can now participate with tandem skydive\footnote{It is not possible to experience speed skydiving through tandem skydiving for non-skydivers.}. However speed skydiving is an advanced level of skydiving as it requires a good command of motion through the freefall and additional procedures in order to complete the jump in a safe manner.  To elaborate on this, we will walk the readers through the whole process of a typical speed skydiving. The first step being an exit from the aircraft at 4000 meters altitude. The second being the action itself, \ie the speed skydiver forms herself into a streamlined shape in order to penetrate the ocean of air towards the altitude of 1700 meter after which the measurement ends. The speed skydiver would then have to try to slow down in order to deploy her parachute before 1000 meters altitude and subsequently fly it down to a designated landing site. The measuring device is then handed over to a team of judges at competitions or the jumpers themselves just take a look at the device for her own record during training. 

No other disciplines in skydiving measures freefall speeds as a part of competitions. Understanding how freefall speeds change with mass, time, surface area as well as air density would enhance our  understanding of downward motions with air resistance--especially in educational settings having speed skydiving as a case study. Our theoretical calculations can directly be compared with the measurements done using the equipment that the athletes utilize.

Note that we use the terminology "freefall" despite the fact that skydivers are not really in a freely falling motion due to atmospheric frictions. Literally, there is no genuine freefall in Earth's atmosphere, not even on Mars.  It was recently shown by Jet Propulsion Laboratory of NASA that rotocraft (Mars helicopter named Ingenuity) could fly in Martian atmosphere despite the fact that pressure of Martian atmosphere is only approximately 1\% of Earth's~\cite{NASAmars}.

\section*{Speed skydiving as a sport discipline}

It is the International Skydiving Committee (ISC\footnote{https://www.fai.org/commission/isc (cited on May 14, 2021).}) within FAI that passes rules for competitions in all disciplines of skydiving.   According to latest regulation, the score of a speed skydiving is the average vertical speed in km/h to the nearest hundredth of a km/h of the fastest 3 seconds, which the competitor achieves within the performance window.  A speed skydiver has approximately 7400 ft for her performance (measured from the minimum exit altitude and breakoff altitude). It is clearly understood that the speed skydiver has to make sure that during a given distance, the best geometrical form is achieved as she has to reach the highest terminal speed.

Measurement of the average speed is done using an electronic equipment based on GPS. Prior to utilization of GPS as an official way of measuring, barometric measurement was used\footnote{In the old system of measurement, the world's record of 601.26 km/h was made by Mr. Henrik Raimer of Sweden at World Championships in IL, USA in 2016~\cite{raimer}. }. Currently one can find rankings in both systems on the website of The International Speed Skydiving Association (ISSA)~\cite{ISSAweb}.

\section{Derivation of terminal speed}

The setup for obtaining the terminal speed is accomplished using Newton's second law of motion. We consider all the forces under consideration acting on the center of gravity of the speed skydiver.  This applies in general to any object falling through atmosphere\footnote{This would be true on Mars as well but may not apply well for other planets whose atmospheres are not gaseous.}. There are only two forces considered, the first being the force due Earth's gravitational field $F = mg$ with $m$ being total mass of the skydiver including her gear (the so-called exit mass) and $g$ being acceleration due to gravity at the location of skydiving. The second one being the atmospheric friction or drag force given by  
\begin{equation}
F_d = kv^2,
\label{eq:drag}
\end{equation}
where  $k = \onehalf \rho C_d A $ with $\rho$ being air density, $C_d$ a dimensionless drag coefficient (body orientation of a skydiver) and $A$ the cross sectional area of the skydiver.  Taking into account the direction of motion (with the positive direction pointing downwards) we get
\begin{equation}
ma = F_{grav} - F_d.
\label{eq:3forces}
\end{equation}
Expressing the LHS of the Eq.(\ref{eq:3forces}) in terms of speed as a function of time we have 
\begin{equation}
\frac{dv}{dt} = g - \frac{kv^2}{m}. 
\label{eq:DE1}
\end{equation}
We have two options, \ie solve  Eq.(\ref{eq:DE1}) for $v$ as a function of $t$ or as a function of a vertical displacement $y$.   

In both cases we will see that the amount of time (and vertical displacement) required to reach a terminal speed from a jump from an aircraft is affected by what is contained in $k$.

\subsection{Terminal speed as a function of displacement/altitude}

In order to solve Eq.(\ref{eq:DE1}) for $v(y)$ we would have to rewrite it using a chain rule:
\begin{equation}
\frac{dv}{dt} = \frac{dv}{dy}\cdot\frac{dy}{dt} = g - \frac{kv^2}{m}, 
\end{equation}
with $\dis \frac{dy}{dt} = v$. Rearranging the equation above we arrive at
\begin{equation}
\frac{\dis v dv}{\dis g - \frac{kv^2}{m}} = dy.
\label{eq:integral1}
\end{equation}
Substituting $ u = v^2$ in Eq.(\ref{eq:integral1}) gives 
\begin{equation}
\frac{du}{2 \paren{ g - \frac{ku}{m}} } = dy,
\end{equation}
which after integration yields
\begin{equation}
-\frac{m}{2k} \ln (mg -ku ) = y + C,
\end{equation}
where $C$ being a constant to be determined. The initial speed of a given speed skydiver before she exits the aircraft is zero, hence 
$\dis C = -\frac{m}{2k} \ln(mg)$. This gives us the vertical displacement 
\begin{equation}
y = -\frac{m}{2k}\ln\paren{1 - \frac{kv^2}{mg}}. 
\label{eq:vertical}
\end{equation}
Solving Eq.(\ref{eq:vertical}) for $v$ yields
\begin{equation}
\label{eq:speeddistance}
\dis v(y) = \sqrt{\dis \frac{mg}{k}\paren{\dis 1 - e^{\dis -2ky/m} }} \, .
\end{equation}
Reinserting $k$ in the equation above we arrive at
\begin{equation}
v(y) = \sqrt{\frac{2mg}{\rho C_d A}} \paren{\dis 1 - e^{\dis \frac{-\rho C_d A y}{m}}}^{1/2}.
\end{equation}
We can readily see that the square-root factor is the terminal speed\footnote{which can easily be obtained from setting the LHS of Eq.(\ref{eq:3forces}) to zero, \ie when the motion is acceleration-free.}, $v_T$. Graphically we can read off when $v_T$ is reached. Mathematically it is when the second term in the parenthesis vanishes, meaning when $y \rightarrow \infty$. In \cite{brazil} the terminal speed is calculated directly from equilibrium, \ie when drag and gravity pull are equal.

\subsection{Terminal speed as a function of time}

To obtain terminal speed as a function of time we rewrite Eq.(\ref{eq:DE1}) as 
\begin{equation}
\dis dt = \frac{\dis dv}{\dis g \paren{ 1 -  kv^2/mg}}.
\end{equation}
Integrating this equation\footnote{With initial conditions $t=0$ and $v=0$ disregarding the speed of the aircraft prior to the exit.} yields
\begin{equation}
t = \frac{1}{g\gamma} \tanh^{-1}(\gamma v)
\end{equation}
where $\gamma = \sqrt{k/mg}$. Reinserting $k$ (from Eq.(\ref{eq:drag})) and $\gamma$ and inverting the expression to obtain the terminal speed as a function of time we arrive at	
\begin{equation}
v(t) = \sqrt{\frac{2mg}{\rho C_d A}} \tanh \paren{\sqrt{\frac{\rho C_d A g}{2m}}t}.
\label{eq:speedtime}
\end{equation}
The expression together with Eq.(\ref{eq:speeddistance}) give us a clear picture of how the terminal speed is attained from an exit points. 

\begin{figure}[htp]
\begin{center}
\includegraphics[scale=0.45]{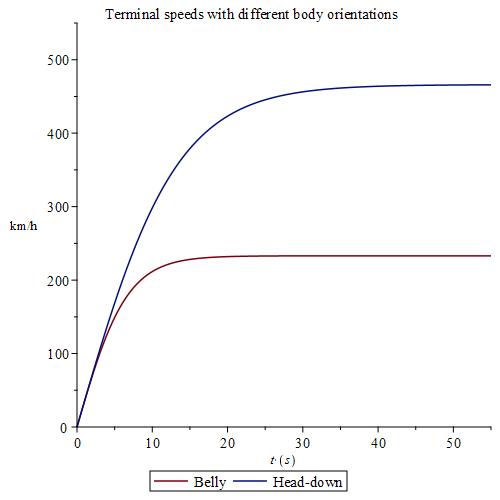} 
\caption{\sl Typical graphs of terminal speeds as a function of time for skydivers in a belly position versus hypothetically a head-down position. In this example we use the author's data, \ie exit mass $m  = 82$ kg. For the belly position we use a drag coefficient of a half-sphere, $C_d = 0.4$ and a cross-sectional area of 0.8 m$^2$. Whereas in the head-down position we assume $C_d = 0.2$ and the cross-sectional area is reduced to 0.4.  In both cases we use $g=9.81$ \; m/s$^2$ and $\rho = 1.23$  kg/m$^3$. It is readily seen that the terminal speeds vary drastically due to body geometry through air. The mass of the skydiver is invariant.}
\label{fig:naritgraphs}
\end{center} 
\end{figure}


\noindent It is natural to assume that when we make a transfer in freefall from a standard belly position to a head-down position (for achieving a greater terminal speed) the only parameters affected by the process are $A$ and $C_d$\footnote{The appropriate drag coefficient for skydiving is rather uncertain. Different values are given in different sources. As a skydiver myself, the belly position is not a plat plane, rather an arch.} as the skydiver cannot reduce or gain her mass during the jump. We will study professional speed skydivers in the next section. It is to be noted that as $m$ increases while $A$ decreases the outcome would be that a skydiver reaches her terminal speed later and the terminal speed becomes greater. This is a common experience by all skydivers (and this is translated into how they learn to move their body during the fall through air resistance of Earth's atmosphere).

\subsubsection{Three other parameters as a function of time?}

It is clear that air density $\rho$, drag coefficient $C_d$ and the cross-sectional area $A$ as appeared in Eq.(\ref{eq:speeddistance}) and Eq.(\ref{eq:speedtime}) of the skydiver for each jump vary. It is however not the purpose of this paper to model each jump, rather we aim to give an overview of the underlying physics as well as comparisons with the current state of development of the sport.

\section{Theoretical models versus reality}

We can argue that air density $\rho$ is a function of altitude (time) for the sake of mathematical modeling. Body position is also a function of time as the speed skydivers need time to adjust their body position so $C_d = C_d (t)$ and $A = A(t)$ but it is nearly impossible to model skydiving in this way as each jump is unique. The air density being a function of altitude (which can be translated into the elapsed time from the exit) is unnecessary unless the jump under consideration is done from a very high altitude.   It is conceivable that no skydivers can dive out of an aircraft into a planned body orientation instantly. They perform the exit in such a way that they gradually transform their body into the head-down and use their arms and legs to control the streamlined flow through the atmospheric frictions. A fully detailed modeling should cover the skydiver's transition from the exit to the final stable body position with $C_d$ and $A$ as functions of time.  We leave this to future investigations.

\subsection*{Mathematical models of speed skydivers}

In private communication with three experienced competitive speed skydivers we are able to estimate their body orientation/geometry in numbers. It it obvious that their total mass, cross-sectional area and drag coefficient play a sensitive role in their terminal speed.  Our method for an approximate model is as follows:  the speed skydivers's best achieved terminal speeds are known, their cross sectional areas are approximated from interviews with them. Since this article is meant to be educational rather than a detailed mathematical modeling, hence we will not stress on the drag coefficients. We will see that $C_d$ varies greatly from the belly position to the speed skydiving position. We utilize Maple 2021~\cite{maple} as a tool for numerical computations and graph plotting.

\begin{figure}[h!]
\begin{center}
\includegraphics[scale=0.3]{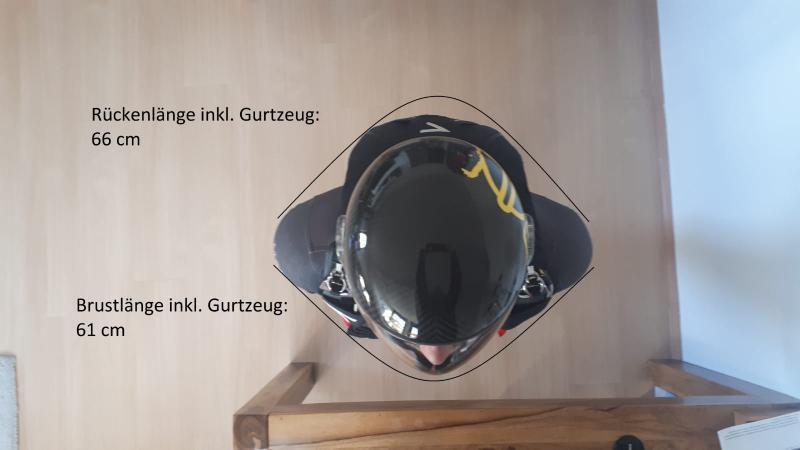} 
\caption{\sl Approximation of Marco Hepp's cross sectional area. Photo: Marco Hepp}
\label{fig:marcohepp}
\end{center} 
\end{figure}

Having these numbers together with other parameters such as air density, acceleration due to Earth's gravity we are able to obtain their $C_d$ through
\begin{equation}
C_d = \frac{2mg}{\rho A v^2_T}
\end{equation} 
The cross-sectional area, $A$, is provided by the skydiver themselves by rough approximation as rectangular area, see Fig.~\ref{fig:marcohepp} as given by Mr. Marco Hepp, who has won German National Championships in Speed Skydiving.

\noindent Values of $\rho$ and $g$ are the standard values used in all case\footnote{Regardless of location of the jumps performed} in this paper. Highest achieved vertical speeds of the studied speed skydivers are retrieved from ISSA's website.   We use the data from the eternal ranking measured by GPS~\cite{ISSA}.  

It is conceivable that two biggest factors that affect the outcome of a speed skydive is $C_d$ and the cross-sectional area $A$\footnote{ A small parachute rig might change the cross-sectional area somewhat.} once in the head-down position. The mass of the skydiver is a fixed number.  The mass of the speed skydivers is, as a matter of fact, one factor affecting the terminal speed. It is clear that the largest the mass, the higher the terminal speed provided everything else remains invariant.  

\begin{figure}[h!]
\begin{center}
\includegraphics[scale=0.4]{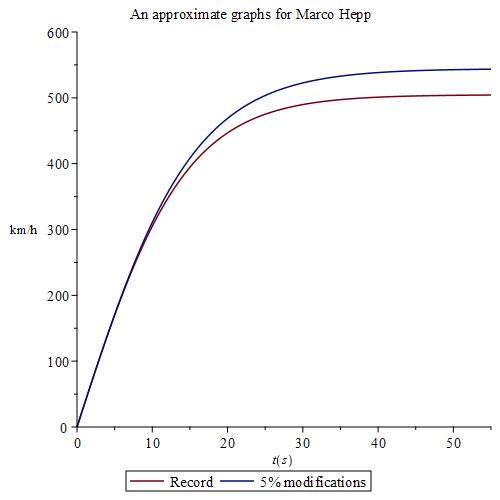} 
\caption{\sl An approximate modeling of terminal speeds for Marco Hepp with his record speed and with 5\% modifications. With the proposed 5 \% modifications the top terminal speed for Marco would increase with 7.7\% reaching in excess of 540 km/h.  }
\label{fig:marcograph}
\end{center} 
\end{figure}

\begin{figure}[hb!]
\begin{center}
\includegraphics[scale=0.4]{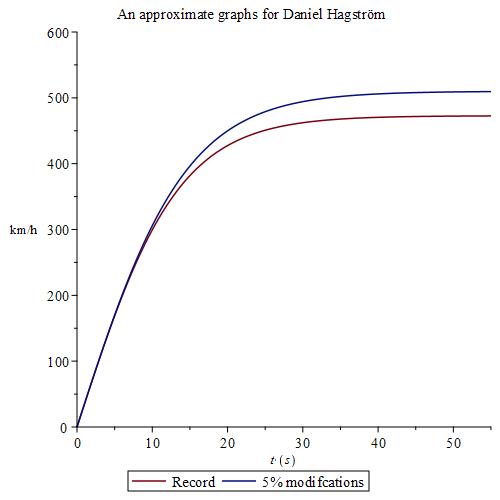} 
\caption{\sl A graph of the freefall speed as a function of time for Daniel Hagström with his record speed and with 5\% modifications. With the modifications, Daniel's theoretical top terminal speed would be above 500 km/h.}
\label{fig:danielgraph}
\end{center} 
\end{figure}

\begin{figure}[htb!]
\begin{center}
\includegraphics[scale=0.4]{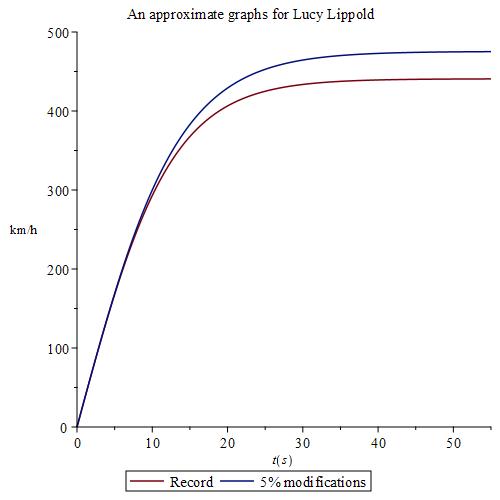} 
\caption{\sl An approximate modeling of terminal speeds for Lucy Lippold with her record speed and with 5\% modifications. With the modifications, Lucy would fly towards the Earth's surface in the vicinity of 475 km/h!}
\label{fig:lucygraph}
\end{center} 
\end{figure}

As a gedankenexperiment we propose theoretical improvements for the three speed skydivers by a so-called 5 \% modifcations meaning that $C_d$ and $A$ are reduced by 5 \% whilst the mass increases by 5 \%. With these modifications we can see how their record terminal speeds improve as shown in Fig.~\ref{fig:marcograph}, \ref{fig:danielgraph}, \ref{fig:lucygraph}.

It is proposed that a standard drag coefficient for speed skydivers is in the vicinity of $0.25$. One might argue that $C_d = 0.2$ is a goal to achieve for anyone wishing to pursue this airsport discipline.

\subsubsection*{Marco Hepp of the German National Team for Skydiving}

Marco Hepp has an exit weight of 84.2 kg with $A = 0.66 \times 0.61$ m$^2$, we get  $C_d = 0.17$ using his top vertical speed of 139.74 m/s.

\begin{table}
\begin{center}
 \begin{tabular}{|l|c|c|c|c|c|}
 \hline 
 Skydiver & $m_{\text{exit}}$ (kg) & $A$  (m$^2$)& $v_{\text max}$ in m/s & $C_d$    \\ 
 \hline 
Lucia Lippold & 73 & 0.30 &   116.02 & 0.26  \\ 
 \hline 
Marco Hepp & 84.2 & 0.40 &   139.74 & 0.17  \\ 
 \hline 
Daniel Hagström & 98 & 0.46 &   131.21 & 0.20 \\ 
 \hline 
 \end{tabular} 
\caption{Data for our approximate modeling of experienced speed skydivers.}
\end{center}
\end{table}

\newpage 

\subsubsection*{Daniel Hagström of the Swedish National Team for Skydiving}

Daniel Hagström has an exit weight of 98 kg with $A = 0.70 \times 0.65$ m$^2$, we get  $C_d = 0.20$ using his top vertical speed of 131.21 m/s.

\subsubsection*{Lucia Lippold of the German National Team for Skydiving} 

Lucia (Lucy) Lippold who won the European Championship with the speed of 417.68 km/h (116.02 m/s) has $C_d = 0.26$ using her exit weight of 73 kg and a cross-sectional area of 0.30 m$^2$.

\newpage

\section{Summary and conclusion}

Skydiving is a physical activity commonly described in physics textbooks at all levels, especially in educational settings. It is a direct demonstration of how Newton's laws of motion are applied for objects moving in atmosphere with air resistance.  When it comes to detailed mathematical modeling of skydiving it is a varying challenge in that it involves many variables which can be considered to be dependent on time or vertical trajectory since the skydiver's body orientation is not fixed ($C_d$ and $A$ as functions of time/altitude can be a subject in its own right\footnote{We should be able find approximations in which $C_d(t)$ and $A(t)$ from the exit altitude be linear functions which can then be inserted into Eq.(\ref{eq:3forces}). Solving it for $v(t)$ or $v(y)$ would directly be more demanding.}). 

Gravity, air resistance, basic aerodynamics and calculus have been the main tools for discussing skydiving from exit to deployment of a parachute to the very end of the activity, \ie the landing with parachute. Activities discussed using the concept of terminal speed and utilization of parachute(s) within physics education include landing of spacecraft returning to Earth as well as the recent landings of spacecraft on Mars~\cite{NASAmars} and Titan--the largest moon of Saturn in 2005~\cite{titan}. A well-known terminology for this is EDL (Entry Descent Landing)~\cite{EDL}. We do not study a parachuting part of speed skydiving in this paper.  In studying/designing EDL, in particular through atmosphere the knowledge $C_d$, object's shape, cross-sectional area, terminal speeds, gravity as well as atmospheric density are necessary. See \eg~\cite{marspaper} for physics of landing on the surface of Mars and other planets. 

Speed skydiving is about the terminal speed which is mainly affected by what can be controlled by the skydiver herself.  Body's mass, gravity and air density are in principle constants for each computation, but $C_d$ and $A$ do play a vital role for each jump. This can be translated into a training procedure for this sport. We have illustrated the underlying physics of speed skydiving, the way the sport is competed and presented some data from world class's speed skydivers. It is proposed that the drag coefficient for speed skydiving is in the vicinity of 0.25. Further investigations of this in wind tunnels with exact measurements and detailed mathematical modeling for each speed skydiver are encouraged.

\subsection*{Acknowledgments}

The author would like to extend sincere thanks to the following persons for feedback and various discussions: Stefan Ericson of University of Skövde, Surachate Kalasin of King Mongkut's University of Technology Thonburi, Gerda-Maria Klosterman-Mace, Marco Hepp, Daniel Hagström and Lucy Lippold.  A special thank to Bob Kucera and Isabella Malmnäs for valuable comments.

\vspace{1cm} 

\end{document}